\documentclass[aps,prl,preprint,amsmath,showpacs,floatfix]{revtex4}
\usepackage{epsfig}
\begin{document}
\preprint{}
\title{Generation of Traveling Surface Plasmon Waves by Free-Electron Impact}
\author{M. V. Bashevoy, F. Jonsson, A. V. Krasavin, N. I. Zheludev}
\affiliation{EPSRC Nanophotonics Portfolio Centre,
  School of Physics and Astronomy, University of Southampton,
  SO17 1BJ United Kingdom}
\author{Y. Chen}
\affiliation{Rutherford Appleton Laboratory Didcot, Oxon, OX11 0QX
  United Kingdom}
\author{M. I. Stockman}
\affiliation{Department of Physics and Astronomy, Georgia State University,
  University Plaza Atlanta, GA 30303-3083 United States}
\date{April 28, 2006}
\begin{abstract}
\noindent
The injection of a beam of free 50 keV electrons into an unstructured gold
surface creates a highly localized source of traveling surface plasmons with
spectrum centered around 1.8 eV. The plasmons were detected by a controlled
decoupling into light with a grating at a distance from the excitation point.
The dominant contribution to the plasmon generation appears to come from the
recombination of d-band holes created by the electron beam excitation.
\end{abstract}
\pacs{}
\maketitle
\noindent
Surface plasmon polaritons (SPPs) are coupled transverse electromagnetic
field and charge density oscillations which propagate along the interface
between a conductor and a dielectric medium. The main feature of the SPPs
that currently attracts exploding attention is that they are strongly
localized, making them favored candidates as information carriers in
applications such as high-density broadband interconnections and signal
processing~\cite{Barnes:2003,Zayats:2005}.

The excitation of SPPs is usually performed by optical means, and
since SPPs do not couple to light illumination at flat
metal-vacuum interface, the energy coupling is achieved using
gratings or prism matching schemes. Surface plasmon waves can also
be generated in corrugated tunnelling
junctions~\cite{Kirtley:1980}. Unfortunately, these techniques do
not easily allow for a high localization of the SPP source, as
being essential for nanophotonic devices. Discontinuities of a
plasmon waveguide, such as a nanoparticle,
nanowire~\cite{Krenn:2003} or nanoscale aperture of a nearfield
optical probe~\cite{Sonnichsen:2000}, may be used for a more
localized launch of plasmon waves. However, these coupling
techniques are cumbersome, and they do not always allow for easy
repositioning of the plasmon source. Ritchie predicted that the
electron bombardment of a metal film could lead to the excitation
of surface plasmons~\cite{Ritchie:1957}, and evidence of this was
later observed in aluminium films by electron energy loss
spectroscopy~\cite{Powell:1959} and by light emission of silver
grating surfaces~\cite{Teng:1967}. Recently an evidence of
propagating surface plasmon modes was observed in the spatial
distribution of optical emission on microscale gold corral under
electron excitation in the scanning tunneling
microscope~\cite{Beversluis:2003}.

In this Letter, we report on the first demonstration of direct
excitation of SPPs by injection of a beam of free electrons on the
\emph{unstructured} metal interface, creating a plasmon source
potentially with nanoscale localization, which may be easily and
dynamically repositioned anywhere in a plasmonic device, for
example in a SPP waveguide.

\begin{figure}[h]
\includegraphics[width=140mm]{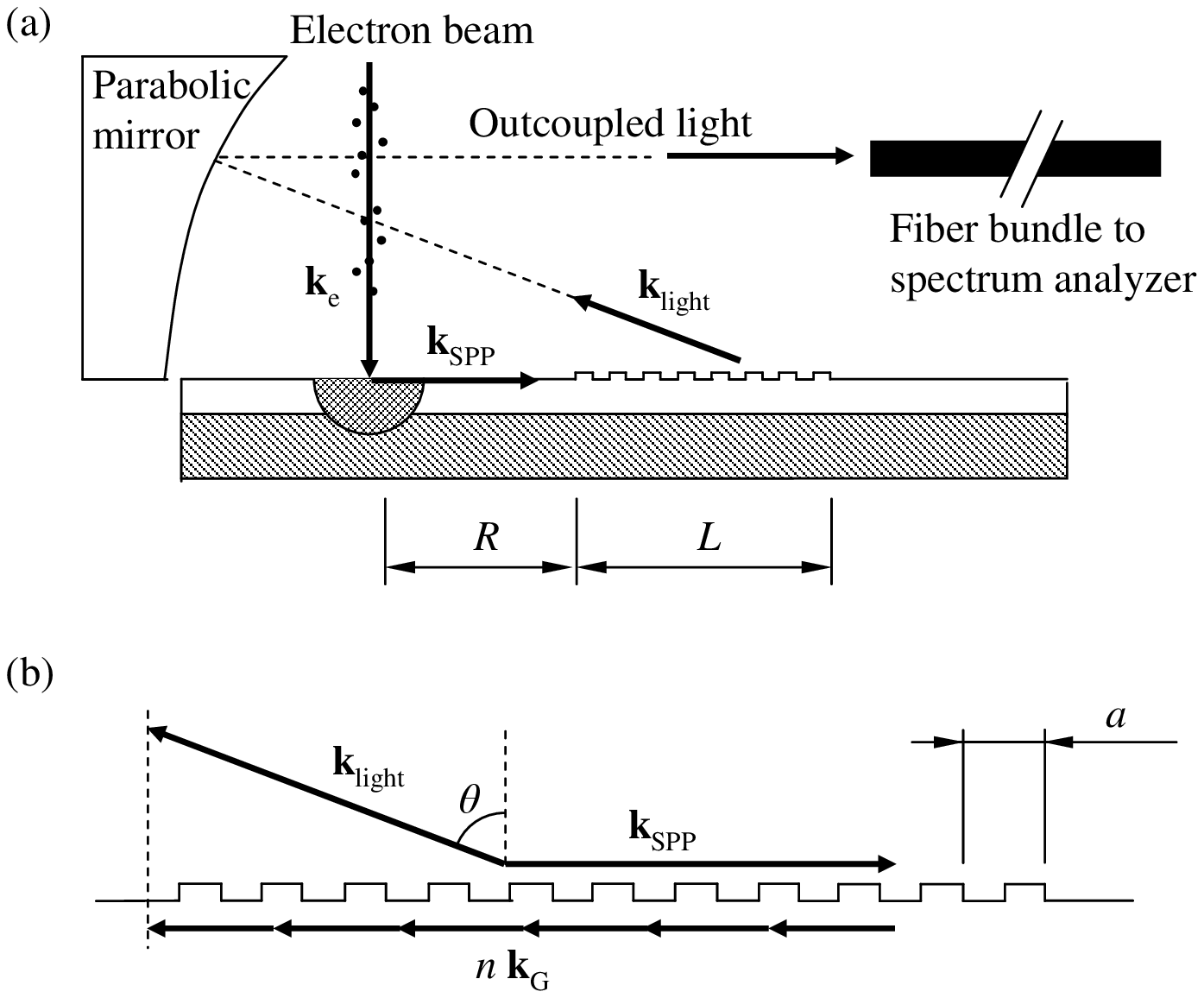}
\caption{\label{fig:1}
  Schematic of (a) the experimental setup for excitation of surface plasmon
  polaritons (SPPs) with wave vector ${\bf k}_{\rm SPP}$ by direct injection
  of a beam of free electrons of wave vector ${\bf k}_{\rm e}$, and their
  decoupling as light by a grating, and (b) the geometry of the wave vector
  matching between SPPs, uncoupled light and the grating.}
\end{figure}

A fast electron penetrating a metal film loose a fraction of its
energy by exciting plasmons on the metal surface, as
illustrated in Fig.~\ref{fig:1}. In our experiments the SPPs were
excited in gold films by a focused electron beam of a scanning
electron microscope. The SPPs were decoupled into light by a
microscopic grating manufactured on the metal surface, with the
uncoupled light being collected by a parabolic metal mirror into
an optical multichannel spectrum analyzer consisting of a
Jobin-Yvon C140 spectrograph and a liquid nitrogen cooled CCD
array, in a geometry as illustrated in Fig.~\ref{fig:1}(a). The
samples were $200\,{\rm nm}$ thick gold films which incorporated
$50\,{\rm nm}$ deep gratings with a period of $4.25\,\mu{\rm m}$.
The gratings were specifically designed for decoupling of light
predominantly in the backward direction.

\begin{figure}[h]
\includegraphics[width=140mm]{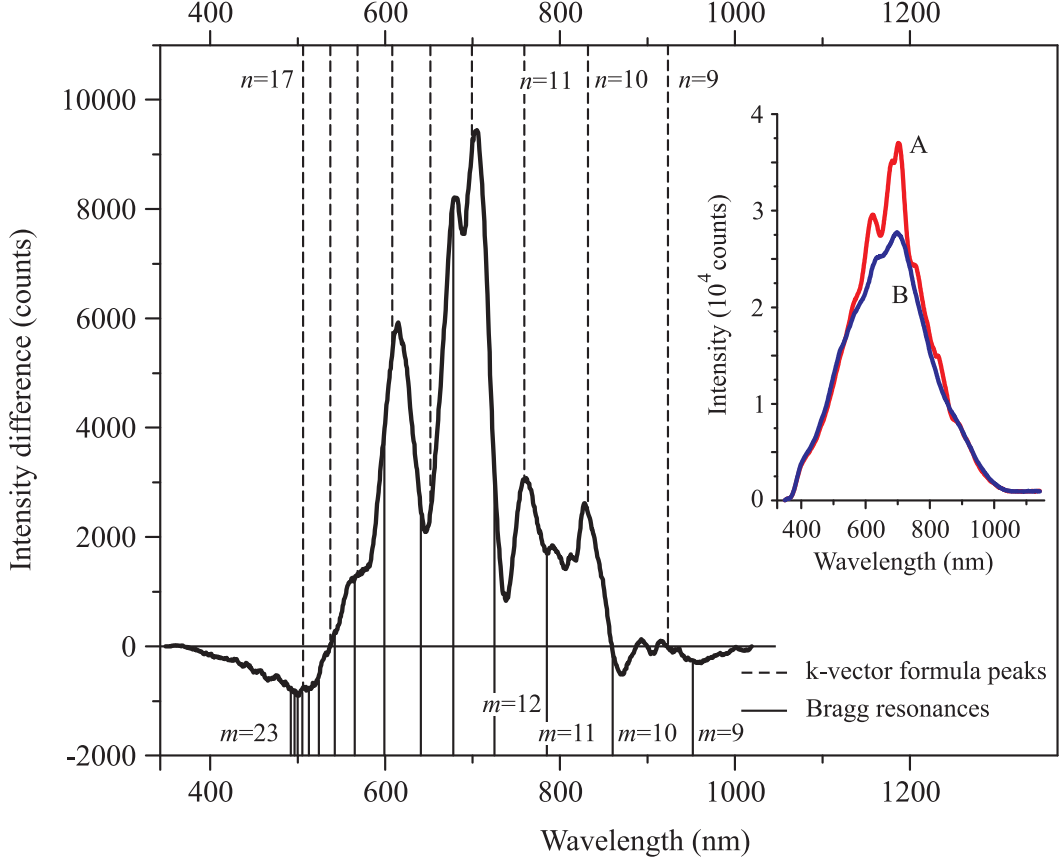}
\caption{\label{fig:2}
 Differential spectrum of light emission of a gold grating of period
 $a=4.25\,\mu{\rm m}$, oriented with its ribs perpendicular to the
 direction of the mirror and excited by a beam of electrons of
 $50\,{\rm keV}$ energy at a beam current of $12\,\mu{\rm A}$.
 The inset shows non-normalized emission spectrum of (A) the grating and
 and (B) the unstructured gold surface.
 The differential spectrum is obtained by subtracting spectrum (B) from
 spectrum (A), hence implying that the differential spectrum only consists
 of the impact of the grating, eliminating transient radiation and
 luminescence of the gold film.}
\end{figure}

The evidence of SPP generation by electron beam excitation was
seen in two series of experiments. In the first series we compared
light emission from the unstructured gold surface with emission
from the gold grating. In this case the experiments were performed
using scanning mode of the microscope, with a scan area of about
$130\times100\,\mu{\rm m}^2$. Optical emission of the unstructured
gold surface was clearly seen in our experiments. This is a
combination of the $d$-band fluorescence, as previously seen in
femtosecond photoluminescent experiments~\cite{Beversluis:2003},
transient radiation of the collapsing dipoles formed by the
electrons approaching the metal surface and their oppositely
charged mirror image, and the fluorescence of any residual
contaminants on the sample. These mechanisms of emission create a
smooth spectrum centered at about $700\,{\rm nm}$, shown as line~B
in the inset of Fig.~\ref{fig:2}. The emission spectrum of
unstructured gold surface was identical to that of the grating
parallel the direction towards the collection mirror. However,
when the grating was oriented with its ribs perpendicular to the
direction to the mirror, the emission was stronger than of the
unstructured gold and its spectrum showed some pronounced
modulations.
Such a difference in detection of emission is explained by the directionality
of the decoupling of the travelling SPPs, which only in the latter case are
decoupled into light directed towards the detection system.

A grating fabricated on a metal surface facilitates decoupling of
light by providing a wave vector mismatch equal to an integer
multiple of the grating vector $k_{\rm G}=2\pi/a$ where $a$ is the
grating period, as illustrated in Fig.~1(b). Only SPPs of certain
frequencies are decoupled by the grating in the direction of the
detector, at an angle of approximately $\theta=70$~degrees, as
determined by the geometry of setup. This decoupling is described
by the kinematic equation
\begin{equation}
  {\rm Re}\{k_{\rm SPP}(\omega)\}-n k_{\rm G}=(\omega/c)\sin(\theta(\omega)),
\label{eq:2}
\end{equation}
where $n$ is a positive integer describing the diffraction order.
In the spectrum shown in Fig.~\ref{fig:2}, the peak wavelengths at which
optimum detection of the decoupled radiation occur are indicated with the
corresponding diffraction orders~$n$, as calculated from Eq.~(\ref{eq:1}).
One can clearly see that peaks in the emission spectrum indeed to a high
degree coincide with the predicted values for efficient SPP emission.
One also can also observe several dips in the emission spectrum and even
wavelengths at which the emission from the grating is less intense than
that of unstructured gold. This is believed to result from the modification
of the plasmon dispersion by the grating at frequencies
$\omega=\omega_{\rm res}$ given by the Bragg condition
\begin{equation}
  m k_G/2={\rm Re}\{k_{\rm SPP}(\omega)\}= \frac{\omega}{c}
    {\rm Re}\bigg\lbrace\bigg(\frac{\varepsilon(\omega)}
               {1+\varepsilon(\omega)}\bigg)^{1/2}\bigg\rbrace,
\label{eq:1}
\end{equation}
where $m$ is a positive integer describing the resonance order.
Here $\varepsilon(\omega)$ is the complex-valued permittivity of gold.
Corresponding wavelengths of Bragg resonance are in Fig.~\ref{fig:2}
presented for various values of $m$, indicating that the SPPs at Bragg
resonance are either badly coupled to light or their generation by the
electron beam is inhibited by the grating.

The second experimental series aimed to demonstrate the generation
of travelling SPP waves on an \emph{unstructured} metal surface,
their propagation and controlled decoupling into light by a
grating. In these experiments we studied the dependence of the
optical emission as function of distance~$R{}$ between the
excitation point and the grating edge. This series of experiments
was performed using spot mode of excitation, with a spot diameter
of about $1\,\mu{\rm m}$. The large spot size was used in order to
allow higher beam current and more intense peaks in decoupled light.
The spectrum of the signal detected by placing the electron beam at
the edge of the grating facing the mirror is essentially the same
as in the scanning mode over the grating, with the only difference
that the negative values in the differential spectrum seen in
Fig.~\ref{fig:2} at about $\lambda=500$, $870$ and $950\,{\rm nm}$
do not appear, corroborating with the idea that they are indeed
relevant to the Bragg frequencies.

\begin{figure}[t]
\includegraphics[width=140mm]{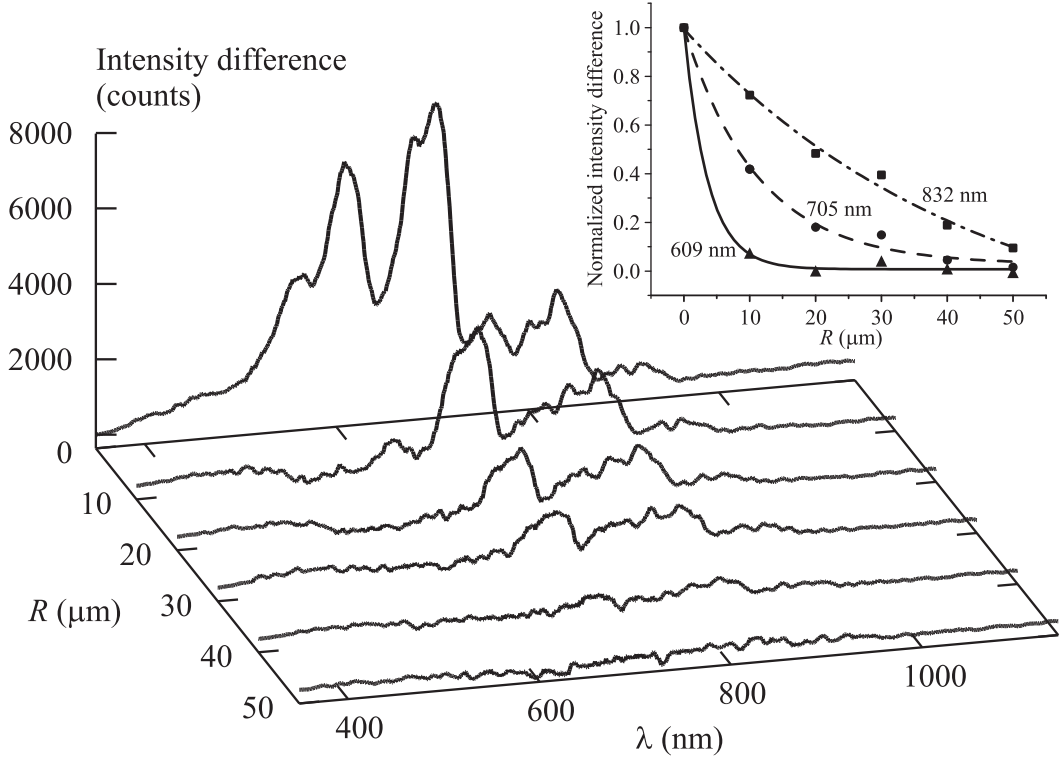}
\caption{\label{fig:3} Decay of the SPPs as function of distance~$R{}$ between
the grating edge and electron injection point.
The differential spectra were obtained by substracting the spectrum sampled
at $R{}=60\,\mu{\rm m}$ from the spectra sampled at shorter distances.
The inset shows normalized intensities of decoupled SPP signal at different
peak wavelengths as function of distance $R{}$ between the edge of the grating
and the point of excitation.}
\end{figure}

As the point of excitation is moved away from the grating, the
spectrum gradually changes and the SPP component of emission
diminishes, as shown in Fig.~\ref{fig:3}. This graph illustrates
that SPP waves corresponding to different parts of the emission
spectrum decay  with different pace (see inset into Fig.~\ref{fig:3}).
For a point-like source, the in-plane plasmon intensity is proportional
to $e^{-R/\xi}/R$.
Due to our detection system, collecting light in a range of in-plane
azimuthal angles, this leads to an $e^{-R/\xi}$ dependence for the decoupled
and detected signal, giving the experimental attenuation lengths
$\xi_{609}=5\,\mu{\rm m}$, $\xi_{705}=11\,\mu{\rm m}$, and
$\xi_{832}=45\,\mu{\rm m}$.
Indeed, plasmons corresponding to the vacuum wavelength of
$609\,{\rm nm}$ shall be strongly attenuated by losses in gold
with damping length rapidly increasing towards the infrared part
of the spectrum. This is also what we find in our experimental
data. However, our experimentally derived energy attenuation lengths
are somewhat shorter than predicted from the formula $\xi=(2|{\rm
Im}\{k_{\rm SPP}\}|)^{-1}$ and the bulk values of the dielectric
coefficient of gold. This is explained by imperfections and
granulation of the gold surface, providing an additional source
for plasmon scattering losses.

All our observations, in particular the wavelength dependent decay
of the emission spectrum with the distance between the excitation
point and the grating, prove that electron beam excitation indeed
provides a source of SPPs. The plasmon emission spectrum largely
correlates with the spectrum of unstructured gold emission as
shown in graph~B of Fig.~\ref{fig:2}, suggesting that the dominant
contribution to SPP generation comes from the recombination of
$d$-band holes created by electron beam
excitation~\cite{Dulkeith:2004}, rather than from direct
scattering of free electrons ~\cite{Ritchie:1957}. Therefore it
appears that plasmon emission spectrum, as the spectrum of
luminescence are strongly connected to the energy separation
between $d$ holes and the Fermi surface near $X$ and $L$ points
roughly at $1.8\,{\rm eV}$ ($\lambda=690\,{\rm nm}$), and $2.4\,{\rm eV}$
($\lambda=516\,{\rm nm}$), correspondingly~\cite{Beversluis:2003}.

From our experimental data, one can obtain an estimate of the order of
magnitude of the total power of the SPP source at the point of excitation.
At the electron beam current of $10\,\mu{\rm A}$, or equivalently
$6\times10^{13}$ electrons per second, we detect plasmon-related photons
at the rate of $3\times10^4\,{\rm s}^{-1}$ across the spectrum.
As a rough estimate of the quantum efficiency of our light decoupling,
collection and detection system, we obtain about $10^{-6}$.
This corresponds to a SPP source with a total power of $10\,{\rm nW}$,
generating $3\times10^{10}$ SPPs per second at the point of excitation.
The corresponding probability for a single electron to excite a SPP is
then $3\times10^{-4}$, which is consistent with
Refs.~\cite{Ritchie:1957,Farrell:1958}.

In conclusion, we have shown that electron beam excitation of an
unstructured gold surface provides a potentially highly localized
source of propagating surface plasmons.  This may be the technique
of choice for creating the high density of plasmons necessary for
demonstrating nonlinear regimes of SPP propagation, and also for
achieving a high density of plasmons in the active media of spaser
applications \cite{Bergman:2003}.

This work was supported by grants from the Engineering and
Physical Sciences Research Council (UK), the Office of Basic
Energy Sciences, U.S. Department of Energy, and the US National
Science Foundation. Stimulating discussions with Javier Garcia de
Abajo and useful references provided by Mathieu Kociak are also
acknowledged.

% \bibliography{article} % when using BibTeX

\vfill\eject

\vfill\eject

\vfill\eject

\vfill

\end{document}